\documentclass[sigconf]{acmart}
\usepackage{booktabs}
\usepackage[utf8]{inputenc}
\usepackage{tabularx}
\usepackage{soul}
\usepackage{booktabs}
\usepackage{array}
\usepackage{pgfplots}
\usepackage{tikz} 
\usetikzlibrary{arrows.meta}
\usetikzlibrary{arrows.meta,positioning,calc}
\pgfplotsset{compat=1.18}
\newcolumntype{Y}{>{\raggedright\arraybackslash}X}
\usepackage{tabularx}
\usepackage{array}
\newcolumntype{Y}{>{\raggedright\arraybackslash}X}
\newcolumntype{L}[1]{>{\raggedright\arraybackslash}p{#1}}
\usepackage{array}
\renewcommand{\arraystretch}{0.9} 
\usepackage{makecell}
\usepackage{placeins}
\usepackage{cellspace}
\usepackage{pifont}

\usepackage{enumitem}
\AtBeginDocument{%
  \providecommand\BibTeX{{%
    \normalfont B\kern-0.5em{\scshape i\kern-0.25em b}\kern-0.8em\TeX}}}

\begin{document}

\author{Aditya Kumar Purohit}
\email{aditya.purohit@cais-research.de}
\affiliation{
  \institution{Center for Advanced Internet Studies}
  \city{Bochum}
  \country{Germany}
}

\author{Hendrik Heuer}
\email{hendrik.heuer@cais-research.de}
\affiliation{
  \institution{Center for Advanced Internet Studies
\& \\ University of Wuppertal}
  \city{Bochum}
  \country{Germany}
}

\copyrightyear{2026}
\acmYear{2026}
\setcopyright{cc}
\setcctype{by}
\acmConference[CHI '26]{Proceedings of the 2026 CHI Conference on Human Factors in Computing Systems}{April 13--17, 2026}{Barcelona, Spain}
\acmBooktitle{Proceedings of the 2026 CHI Conference on Human Factors in Computing Systems (CHI '26), April 13--17, 2026, Barcelona, Spain}
\acmDOI{10.1145/3772318.3791763}
\acmISBN{979-8-4007-2278-3/2026/04}

\title[Conditional Companion: LLMs \& Mental Health]{
A Conditional Companion: Lived Experiences of People with Mental Health Disorders Using LLMs}

\begin{abstract}
Large Language Models (LLMs) are increasingly used for mental health support, yet little is known about how people with mental
health challenges engage with them, how they evaluate their usefulness, and what design opportunities they envision. We conducted 20 semi-structured interviews with people in the UK who live with mental health conditions and have used LLMs for mental health support. Through reflexive thematic analysis, we found that participants engaged with LLMs in conditional and situational ways: for immediacy, the desire for non-judgement, self-paced disclosure, cognitive reframing, and relational engagement. Simultaneously, participants articulated clear boundaries informed by prior therapeutic experience: LLMs were effective for mild-to-moderate distress but inadequate for crises, trauma, and complex social-emotional situations. We contribute empirical insights into the lived use of LLMs for mental health, highlight boundary-setting as central to their safe role, and propose design and governance directions for embedding them responsibly within care ecosystem.
\end{abstract}



\maketitle

\section{Introduction}\label{sec:introduction}

The global demand for mental health support is increasing sharply. By 2030, depression is projected to become the third leading cause of disability and premature mortality in developing countries and the second in middle-income nations~\cite{mathers2006projections}. Yet access to care remains critically limited. In low-income regions, there is, on average, only one mental health professional per 100,000 people, compared to 60 per 100,000 in high-income countries~\cite{WHO2022MentalHealth}. Even in high-resource contexts, shortages contribute to long waiting times, while the financial costs of treatment further exacerbate barriers to care~\cite{Garrett2025MentalHealth}.

Recent advances in artificial intelligence, particularly the emergence of Large Language Models (hereafter, LLMs), have been proposed as a means to address these gaps~\cite{Rosteck2025,Sucharat2024}. Their capacity for immediate, around-the-clock conversation makes them appealing in contexts where professional assistance is limited. In conflict-affected areas such as Lebanon, where mental health professionals are rare, LLMs are already being used to deliver support~\cite{Shmais2025ChatGPTLebanon}. Early studies suggest that they may have therapeutic potential. \citet{Sucharat2024} found that patients positively evaluated the empathetic and supportive qualities of LLM-based interactions, while a recent assessment by psychologists concluded that LLMs show strong potential as therapeutic tools, particularly for their skill in empathetic engagement and conversational adaptability~\cite{BerrezuetaGuzman2024}.

At the same time, concerns regarding their limitations are mounting. Media reports describe users who initially found comfort in ChatGPT but later realized that only a human therapist could provide the depth of emotional connection required for recovery~\cite{Shmais2025ChatGPTLebanon}. Regulators in the United States have begun to act, with Illinois passing legislation prohibiting the use of artificial intelligence in therapy services~\cite{IDFPR2025}. Other accounts highlight troubling cases in which AI systems failed to detect suicidal intent or even reinforced harmful ideation~\cite{stanford2025ai-mental-health}. Additionally, recent research reports that LLMs are prone to hallucinations and its tendency to flatter can be detrimental by reinforcing incorrect beliefs in individuals susceptible to psychosis.~\cite{psyhosis2025,stigmaLLMs2025}

This tension between the evidence of promise and harm motivates our research. Despite the growing interest in AI-mediated support, little is known about why individuals turn to tools such as ChatGPT for emotional assistance, the contexts in which they use them, the boundaries they draw around their appropriate use, and the opportunities they envision for safer and more effective systems. Existing research has largely focused on technical capabilities, leaving users' lived experiences, circumstances, and subtle perceptions underexplored.

To address this gap, we conducted 20 semi-structured interviews with people in the UK living with mental health challenges (henceforth, PMHC) who had used LLMs for support. We posed three questions: \textbf{RQ1}. In what ways and under what circumstances do individuals with mental health challenges engage with LLMs for support? \textbf{RQ2}. What are the boundaries of LLM-mediated mental health support relative to human care ? \textbf{RQ3}. What design considerations emerge from user experiences with LLMs in relation to therapy, self-management, and accessibility ?

Our findings show that PMHC were motivated to use LLMs not only for their immediacy and non-judgmental stance but also for the sense of control they afforded over conversation pacing, their ability to support cognitive structuring and reframing, and the relational engagement they provided to patients. At the same time, participants drew clear boundaries around the role of LLMs, shaped by their prior therapeutic experiences. They considered LLMs useful for managing mild to moderate concerns, such as everyday worries, intrusive thoughts, or loneliness, but inadequate for severe situations involving trauma, suicidal ideation, or socially complex challenges that require human judgment and care. Moreover, our research reveals that the participants frequently took it upon themselves to protect their privacy and manage their interactions with LLMs. This behavior prompts important questions about whether the responsibility for safety should lie with individuals or be the duty of companies and regulatory bodies. This paper contributes to HCI community by advancing human-centered understanding of LLM use in high-stakes mental health contexts. \\
\begin{enumerate}[itemsep=0.5ex, topsep=0pt, parsep=0pt, partopsep=0pt]

   \item ~\textbf{Empirical grounding of LLM use as situated care work, not passive adoption}. We provide rich empirical evidence showing how people with mental health challenges adopt LLMs through ongoing sense-making, risk assessment, and self-regulation. This extends prior HCI work on AI adoption by re-framing LLM use as situated care work, challenging dominant techno-optimist and paternalistic assumptions in Human-Centered AI research.\\

  \item ~\textbf{Boundary-setting as a form of user agency and reflective AI engagement}. We introduce an analytic account of how users actively establish and revise boundaries around LLM use based on therapeutic experience and self-reflection. This reframes boundary-setting, often treated as misuse or limitation in AI systems, as a form of competent, reflexive engagement. Moreover, it offers a conceptual lens for CHI researchers studying responsible AI use beyond healthcare.\\

  \item ~\textbf{Design-oriented insights for positioning LLMs within care ecosystems}. We derive design and governance implications that articulate how LLMs can be positioned as complementary, transparent tools within existing care ecosystems rather than replacements for professional care. These insights extend CHI design knowledge for trustworthy AI by grounding system design in users' lived reasoning about safety, responsibility, and trust in high-stakes domains.
 
\end{enumerate}

\paragraph{\textbf{Scope}} This study focuses on how individuals engage with general-purpose large language models (e.g., ChatGPT, Claude, Gemini, Grok) for everyday mental health support. We did not restrict or screen for the exclusive use of general-purpose models; instead, we sought to capture the real-world mix of tools that people naturally turn to. Our aim was not to evaluate clinical or medically approved AI systems. We aimed to understand how people use widely available LLMs in their personal mental health routines.

\section{Background and Related Work}
\label{sec:related}

We organize this review around two strands of work that directly inform our study. The first concerns chatbots and therapy apps, which represent an earlier generation of conversational agents for mental health support (see section ~\ref{sec:beforegpt}). The second considers the emerging role of LLMs in this space (see section~\ref{sec:stateofllm}).
Together, these strands of literature show both the lineage of digital mental health interventions and the novel challenges introduced by LLMs, which motivates our focus on the lived experiences of their everyday use and boundaries. This further sets the stage for understanding the evolution of digital mental health tools and their implications for self-care.

\subsection{Chatbots and Mental Health Self-care}
\label{sec:beforegpt}

Self-care is an important concern for people with mental health challenges (PMHC), as access to professional support is often constrained by systemic barriers, such as provider shortages, long waiting lists, high costs, and gaps in insurance coverage ~\cite{Krausz2018AccessibleAC}. Alongside other methods, such as meditation and exercise, many individuals have turned to chatbots as a support tool ~\cite{trustchatbot2023,Vaidyam2019Chatbots}. Notably, approximately one in three mobile health applications are designed to provide mental health support ~\cite{Anthes2016MentalHealthApp}. Scholarship on chatbots for mental health is both long-standing and wide-ranging. Early explorations can be traced back to Joseph Weizenbaum's ELIZA program in 1966 ~\cite{Weizenbaum1966}, which imitated Rogerian psychotherapists. Crucially, this system was conceived as a testbed for natural language interaction, not as a bonafide therapeutic instrument. Yet it helped inaugurate the use of conversational agents in psychological settings. Weizenbaum was nonetheless disturbed when users began ascribing genuine understanding and empathy to ELIZA. His critique of the uncritical application of computers to sensitive human domains ultimately turned him into a self-proclaimed ``heretic of computer science'', challenging dominant narratives about the limitless promise of artificial intelligence ~\cite{Weizenbaum1976}.

Before the advent of LLMs, most mental health chatbots were built on scripted or decision tree architectures ~\cite{chatbotsinmentalhealth2019}. These constrained systems are considered less error-prone because they do not need to create new responses; instead their responses often feel repetitive and limited in scope ~\cite{chatbotsinmentalhealth2019}. Examples include IDEABot ~\cite{Viduani2023IDEABot} and Tess ~\cite{Joerin2019Tess}, which delivered structured interventions but offered little flexibility. To address these shortcomings, retrieval-based systems have incorporated techniques such as keyword matching and machine learning to select the most relevant reply from a fixed repository ~\cite{Yuan2025}. A prominent example is Woebot ~\cite{Darcy2021HumanLevelBonds}, which delivers cognitive behavioral therapy (hereafter, CBT) ~\cite{Butler2006}) through short daily dialogues and mood-tracking. In a controlled trial, Woebot significantly reduced depressive symptoms over a two-week period, and participants described it as empathetic despite its repetitive style ~\cite{Fitzpatrick2017}. Similarly, Wysa, an AI-driven chatbot, showed that users with higher engagement reported greater improvements in depressive symptoms compared to those with lower engagement. Participants valued Wysa's accessibility and non-judgmental tone, though they noted the absence of a genuine human connection as a limitation ~\cite{Inkster2018Wysa}. These systems demonstrate that chatbots can deliver short-term benefits, but their depth and relational qualities are limited.

Other virtual agent systems have sought to create richer therapeutic interactions. Help4Mood, for instance, combines monitoring of the self with CBT components and shows moderate acceptability and potential efficacy in supporting depression treatment ~\cite{Burton2016}. However, engagement varied; the majority of participants did not use it consistently, and some characterized the interactions as repetitive. Nevertheless, regular users demonstrated a substantial improvement on a validated scale. The 3MR2 system ~\cite{Tielman2017TherapyPTSD} employs an ontology-based questioning framework tailored for PTSD therapy. While innovative, patients often found its therapeutic structure overly rigid, and it was deemed unsafe for autonomous use by individuals with a history of suicidality, severe depression, or substance use. Together, these examples illustrate attempts to move beyond simple CBT while also revealing enduring problems of engagement, rigidity, and safety.

Beyond individual evaluations, reviews of chatbot interventions have synthesized these recurring limitations. A scoping review by ~\citet{Arfan2023} identified a lack of high-quality, long-term evaluations and highlighted the scarcity of in-depth qualitative accounts. Mayor's review ~\cite{Mayor2025} confirmed these issues by reporting that users frequently described chatbot interactions as repetitive, lacking empathy, and unable to handle subtle input. Moreover, most studies have focused on feasibility and acceptability rather than on engagement patterns, relationship building, or the longer-term impact of chatbot interactions on users' mental health trajectories.

Overall, this literature shows both the promise and enduring limitations of early conversational agents for mental health support. They highlight common affordances such as accessibility and non-judgmental engagement, but also recurrent challenges, including repetitiveness, shallow empathy, and limited therapeutic scope. Importantly, existing reviews emphasize the scarcity of qualitative accounts of everyday use. These insights informed our study design by motivating us to examine whether and how such patterns carry over into people's engagement with LLMs and shaped our interview questions around everyday motivations (RQ1) and perceived boundaries of LLM-mediated mental health support (RQ2).

\subsection{LLMs in Mental Health Care}
\label{sec:stateofllm}

Interest in LLMs for mental healthcare has surged in recent years~\cite{WANG2025}. Unlike earlier chatbots that depended on pre-written, topic-specific scripts (see Section~\ref{sec:beforegpt}), LLMs can generate flexible, open-ended responses on sensitive topics related to well-being. This technical shift has spurred diverse research on LLM-assisted mental health support. In this subsection, we organize prior work into three parts: (1) fine-tuning LLMs for mental health, (2) using LLMs for screening and risk assessment, and (3) assessing LLMs as supportive counseling agents. Each part reveals specific gaps that motivate our focus on how individuals with mental health challenges engage with LLMs for support (RQ1), how they understand the boundaries of LLM-mediated support relative to human care (RQ2), and what design considerations emerge from their experiences around therapy, self-management, and accessibility (RQ3). 

\paragraph{Fine-tuning LLMs for mental health.}
One strand of research fine-tunes LLMs on mental health datasets to improve diagnostic or counseling performance.~\citet{Xu2024MentalLLM} fine-tune ``Mental-Alpaca'' and ``Mental-FLAN-T5'' and report higher balanced accuracy than larger models such as GPT-3.5 and GPT-4 on mental-health benchmarks~\cite{Xu2024MentalLLM}. Similarly, MentaLLaMA, a fine-tuned LLaMA2 model trained on expert-annotated social media data, approaches state-of-the-art performance while generating human-readable rationales for its classifications ~\cite{meantalllama2025}. MindWatch uses transformer-based models to detect suicidal ideation and deliver personalized psychoeducation with high accuracy ~\cite{Bhaumik2023MindWatch}. Other work customizes pre-trained LLMs by integrating additional signals. For instance, ~\citet{Poorvesh2025} combine stress predictions from wearable data with a pre-trained LLM to provide stress-related psychotherapy, and ~\citet{Abubakar2024RLChatbotMentalHealth} fine-tune LLaMA to generate emotion-sensitive counseling responses that resemble human conversational style.

Building on these developments,~\citet{Maurya2025LightweightLLMsMentalHealth} compare compact models such as BART-base, small T5, and small FLAN-T5 fine-tuned on public mental-health datasets and show that lightweight LLMs can also produce emotionally supportive counseling responses, highlighting their promise for scalable support. Overall, this strand demonstrates strong technical potential for domain-specific and multimodal fine-tuning. Yet evaluations remain largely restricted to benchmarks, short synthetic scenarios, or crowd workers. Little attention is paid to how people with ongoing mental health challenges weave such models into everyday self-management. This missing everyday and clinical validation motivates our focus on how individuals with mental health challenges engage with LLMs for support in their daily lives (RQ1).

\paragraph{LLMs for screening and risk assessment.}
The second strand evaluates LLMs as tools for screening and diagnosis in clinical workflows. ~\citet{Xu2025} assess the diagnostic abilities of 15 advanced LLMs in a Chinese context and show that DeepSeek-R1, QwQ, and GPT-4.1 outperform other models on knowledge accuracy and diagnostic effectiveness, while also emphasizing persistent hallucination risks and the necessity of professional oversight ~\cite{Xu2025}. To address limitations of traditional questionnaires, where respondents may hesitate to acknowledge mental health difficulties, clinicians have long used open-ended techniques such as the Sentence Completion Test (SCT) ~\cite{SCT2010,Lallemand2022}. However, scoring SCT responses is time-consuming. ~\citet{Huang2025} explore AI agents to automatically score SCT responses, finding that agents can match expert accuracy and reliably flag invalid or misleading answers~\cite{Huang2025}. Similarly,~\citet{Sikström2025} evaluate an AI assistant for clinical interviews. They report that it is perceived as highly empathetic and supportive and can identify true cases more consistently than standard mental-health questionnaires, which are prone to human over- and under-diagnosis ~\cite{Sikström2025}. This strand suggests that LLMs may augment screening and risk-assessment workflows by improving consistency, efficiency, and subjective experiences of assessment. At the same time, it surfaces concerns about diagnostic authority, hallucinations, and the emotional safety of letting automated systems judge people's mental health. These tensions motivate our examination of how participants perceive the boundaries and risks of LLM-mediated support relative to human care when they informally ``self-screen'' or sense-make about their mental state outside clinical settings (RQ2).

\paragraph{LLMs as supportive counseling agents.}
The third line of work examines LLMs as conversational partners that provide therapeutic or emotional support.~\citet{liu2023chatcounselor} introduce ChatCounselor, trained on 260 real-world counseling recordings, and show that it outperforms open-source baselines and produces more interactive responses than the list-style outputs often associated with ChatGPT. Building on this system-building work,~\citet{aditilangchain2024} propose MindGuide, a proof-of-concept LangChain-based chatbot that offers proactive support for depression, anxiety, and suicidal thoughts, although it has not yet undergone clinical validation.

Recent studies shift toward relational and affective outcomes. ~\citet{Dong2025} evaluate GPT-4's capacity for emotional support and find that, in its default form, it performs worse than human counselors, but with targeted prompt engineering, it can shift its responses closer to human quality in specific scenarios. A study of AI companions reports that they can alleviate loneliness as effectively as conversations with other people and more than passive activities such as watching online videos ~\cite{Loneliness2025}. Typing Cure provides an in-depth qualitative account of how people experience LLM chatbots for mental health support, showing that these systems can embody therapeutic values, invite rich disclosure, become part of people's coping toolkit ~\cite{typingcure2025}. Related work examining Reddit posts about LLM use in mental-health settings shows that users value LLMs as an always-available, non-judgmental space, but remain concerned about potential misinformation in health-related advice~\cite{redditmh2025}.

This strand highlights the relational promise of LLMs and begins to document how they feel from the user's perspective, but it leaves several gaps. Much of this research emphasizes specialized chatbots or individual platforms, pays limited attention to people with diagnosed mental health conditions, and offers only partial theorization of how users distinguish between “everyday support” and situations that should be handed off to human care. Typing Cure offers a crucial qualitative anchor, yet it treats LLM chatbots primarily as therapeutic interlocutors and pays less attention to how users position general-purpose LLMs within broader care ecologies, including offline services and personal coping strategies ~\cite{typingcure2025}. Our study extends this strand by focusing specifically on people with mental health challenges in the UK, mapping how they appropriate general-purpose LLMs for everyday regulation, where they locate the limits of LLM-mediated support relative to human care, and what kinds of relational and design qualities they wish these tools had (RQ1–RQ3).

Across these three strands, the literature unfolds a rapidly expanding landscape of LLM applications in mental health from fine-tuned diagnostic models and screening tools to conversational agents that can feel empathetic or companion-like. Yet, the systems described are often evaluated on controlled datasets, short scripted scenarios, or general populations. They rarely examine how individuals with ongoing mental health challenges incorporate general-purpose LLMs into their everyday lives, how they negotiate safety and appropriateness relative to human care, or how design decisions shape therapy, self-management, and accessibility. Given the rapid release cycles of models from OpenAI, Google, Anthropic, xAI, and others, these lived-experience questions are pressing.

\section{Method}\label{methods}
To answer our research questions, we conducted a qualitative study combining semi-structured interviews and a survey of current users of large language models (e.g., ChatGPT, Google Gemini, Microsoft Copilot) for mental health support. We opted for semi-structured interviews because previous studies indicate they are effective in capturing people's thoughts, beliefs, and experiences ~\cite{tisdell2025qualitative,interviewmethod2019}. This method is commonly used by health services researchers to understand the real motivations and experiences of subjects ~\cite{interviewmethod2019}. The interviews explored motivations for adoption, contexts and patterns of use, perceived therapeutic value, and prospective design opportunities, generating in-depth accounts of lived experience with LLM-supported mental health practices.

\subsection{Ethical Consideration}

This study was reviewed by the responsible authorities at the first authors' institution and received IRB-equivalent approval. Prior to the semi-structured interviews, all participants provided informed consent. The consent process detailed the study's purpose, procedures, potential risks and the participants' right to withdraw at any time without penalty. To ensure confidentiality and adhere to ethical standards, all potentially identifying information has been removed from the qualitative data presented in the results.

\subsection{Participant Screening and Survey}

We implemented a two-step screening process for our participants. In phase one, we recruited the participants from the database of Prolific. Prolific allows researchers to select participants based on mental health conditions like anxiety, depression, OCD, or PTSD, as well as those who have used LLMs for mental health support. Prolific ensures legitimacy by conducting more than 50 distinct checks, which involve verifying IDs and eliminating automated accounts ~\cite{HaycraftMee2025}.

In the second phase, a survey was sent to those who agreed to participate and were pre-screened by Prolific. The survey in the second phase included further filters with questions like, ``Which of the following mental health conditions have you been diagnosed with by a medical professional?'' The options encompassed the entire spectrum of depression and anxiety disorders, along with more severe conditions such as schizophrenia, bipolar disorder, and dissociative disorders. Participants with diagnoses like schizophrenia, bipolar disorder, schizoaffective disorder, and substance/medication-induced psychotic disorder were deliberately excluded, as these conditions are more severe and may involve episodes of losing touch with reality, potentially affecting the reliability and comparability of the results ~\cite{Wong2018}. Since this study involved the use of LLMs, participants were also asked, ``Which of the following have you used for mental health support or therapy in the past week?'' This question included popular options like ChatGPT, Gemini, Grok, Deepseek, and allowed participants to specify any other LLMs they might be using.

\subsection{Interviews}
We conducted semi-structured, retrospective interviews with people experiencing mental health challenges who had already used LLMs for mental health support. Our goals were to understand (a) their motivations for turning to LLMs, (b) how and when these tools were integrated into daily life, and (c) how they experienced and evaluated the interactions in comparison with mental health counselors. This exploratory approach aligns with prior work seeking conceptual insight into lived experience~\cite{agapie2022longitudinal}. Using a semi-structured format allowed us to cover key topics consistently while adaptively following up on participant responses. By recruiting individuals diagnosed with mental health challenges who were active users of LLMs for support, we gathered reflections on genuine usage episodes and their context-specific motivations.

\subsection{Participants}
Participants ($n=20$) were recruited via Prolific, selecting individuals who self-reported a diagnosis of a mental health challenge such as anxiety, depression, OCD, or PTSD and used LLMs to help them in their mental health journey. The sample included 20 participants aged 21 to 62 years ($M=41.9$, $SD=12.1$); 11 identified as male (55\%) and 9 as female (45\%). All participants were residents of the United Kingdom at the time of the interviews. Interviews were conducted in English via Microsoft Teams, with video and audio recorded upon participant consent. Table~\ref{tab:participants} provides an overview of the participants.

Participants reported a range of mental health conditions, including depression, anxiety disorders, obsessive-compulsive disorder (OCD), post-traumatic stress disorder (PTSD) and attention-deficit/hyperactivity disorder (ADHD). Several participants described multiple concurrent diagnoses (e.g., depression with anxiety and ADHD). The sample therefore reflects a variety of ages, genders, and clinical profiles, spanning early adulthood to later life, and includes both single and co-morbid conditions.

\renewcommand{\arraystretch}{1.15}

\begin{table}[t]
     \footnotesize
    \setlength{\tabcolsep}{3pt} 
    \small
    \caption{Participant demographics, reported mental health and neurodivergent conditions, LLMs used for mental health support, and frequency of use. Frequency categories: High ($\geq$ once/day), Medium (1--6 times/week), Low ($<$ once/week).}
    \label{tab:participants}

    \begin{tabularx}{\columnwidth}{
        l l c Y Y Y c
    }
        \toprule
        \textbf{ID} &
        \textbf{Sex} &
        \textbf{Age} &
        \textbf{Mental Health Dx} &
        \textbf{Neuro-divergent Dx} &
        \textbf{LLM(s) Used for Support} &
        \textbf{Freq.} \\
        \midrule
        P1  & Man   & 23 & Depression; OCD                & --   & ChatGPT              & High   \\
        P2  & Man   & 21 & Depression; Anxiety            & --   & ChatGPT, Grok        & Medium \\
        P3  & Man   & 44 & Depression; Anxiety            & --   & Gemini; ChatGPT      & High   \\
        P4  & Man   & 36 & Depression                     & --   & ChatGPT              & High   \\
        P5  & Man   & 55 & Depression; Anxiety            & ADHD & ChatGPT              & High   \\
        P6  & Man   & 47 & Depression                     & ADHD & ChatGPT; Gemini      & High   \\
        P7  & Man   & 50 & Depression; Anxiety            & --   & ChatGPT              & High   \\
        P8  & Man   & 37 & Depression                     & --   & ChatGPT              & Medium \\
        P9  & Man   & 52 & Anxiety                        & ADHD & Copilot; ChatGPT     & High   \\
        P10 & Man   & 53 & Anxiety                        & --   & ChatGPT              & High   \\
        P11 & Man   & 46 & Anxiety                        & ASD  & Copilot              & Medium \\
        P12 & Woman & 21 & Depression; Anxiety; PTSD      & ADHD & ChatGPT              & High   \\
        P13 & Woman & 31 & Depression; Anxiety            & ADHD & ChatGPT              & High   \\
        P14 & Woman & 55 & PTSD                           & --   & ChatGPT              & High   \\
        P15 & Woman & 51 & Depression; Anxiety; PTSD      & --   & ChatGPT              & High   \\
        P16 & Woman & 32 & Depression                     & --   & ChatGPT              & High   \\
        P17 & Woman & 62 & Depression; Anxiety; PTSD      & --   & ChatGPT              & High   \\
        P18 & Woman & 35 & Anxiety                        & --   & ChatGPT              & Medium \\
        P19 & Woman & 50 & Depression; Anxiety            & --   & ChatGPT              & High
        \\
        P20 & Woman & 37 & Depression; Anxiety            & --   & Gemini              & High 
        \\
        \bottomrule
    \end{tabularx}
\end{table}

\subsection{Interview Protocol}

The interviews consisted of four main parts, in addition to asking participants basic demographic questions. 

The first part of the process involved thanking the participants and explaining the aims of the study, i.e., learning about their experiences with AI-based tools for mental health support. After gaining permission to record the interview, participants were reassured that their information would remain confidential and that their involvement was entirely voluntary. They were also informed that they could skip any question or end the interview at any time if they felt uncomfortable, while still receiving partial compensation. Furthermore, they were advised that they could reach out to a mental health helpline in the UK by dialing 111 or 999 or inform the interviewer immediately.

The second part, background and usage context, invited participants to reflect on their mental health journey and the support systems they had engaged with, such as therapy, medication, peer support, or digital tools. We then asked how they first encountered AI chatbots (e.g., ChatGPT, Gemini) and what motivated them to try such tools. This section helped situate LLM engagement within participants' broader care-seeking practices, as well as the life circumstances and emotional states that prompted their initial use.

The third part, experiences with LLMs, focused on how and why participants turned to LLMs, what kinds of conversations they typically had, and which symptoms or situations they found the tools useful for (e.g., anxiety spirals, intrusive thoughts, reassurance). We asked participants to describe recent interactions, how they felt before and after using the LLMs, and to reflect on what they perceived as helpful or unhelpful about the responses. This section also included questions about boundaries of use, for instance, moments when they chose not to use LLMs, concerns around privacy, or situations where the LLMs' responses felt generic, unsafe, or emotionally limited.

The fourth part, evaluation and reflection, encouraged participants to take a broader view of LLMs in relation to their care. We first asked participants with prior therapy experience to compare LLMs interactions with human counseling, noting perceived strengths, weaknesses, and relational qualities of each. We then invited them to reflect on how LLMs could be redesigned to better meet their needs, including desired features, interaction styles, or safeguards. This dual focus provided insights into both the comparative value of LLMs and user informed design opportunities within the broader ecosystem of mental health support. Each interview concluded with an open reflection, giving participants the chance to share anything not covered and to debrief after discussing sensitive topics.

\subsection{Qualitative Analysis}

Interviews ranged in length from a minimum of 47m 5s to a maximum of 1h 14m 44s. The mean duration was 59m 39s ($M=59.6$, $SD=5.5$), with a total of approximately 19 hours and 53 minutes of recorded material across all 20 interviews. Audio recordings were transcribed verbatim, with identifying information removed. We applied reflexive thematic analysis~\cite{Braun2006} in six iterative phases. The first author conducted an in-depth qualitative analysis by iteratively reading the data and developing detailed codes capturing participants' experiences. These codes were reviewed and discussed in weekly analytic meetings with the second author, through which they were progressively refined and consolidated. Through iterative processes of organizing, splitting, and merging codes, we developed higher-level categories and themes, along with related subthemes. Ongoing discussions during these meetings supported the refinement and redefinition of codes and themes, surfacing underlying assumptions, alternative interpretations, and latent meanings within the data, and ultimately leading to analytic consensus. Guided by an interpretivist approach, we regarded themes as insights co-constructed from both participant and researcher perspectives.

As researchers, we recognize that our social and cultural backgrounds have influenced our interaction with the data. The first author hails from India, while the second author is from Germany. Both authors also bring insights shaped by their education and life experiences in Western and Northern Europe, the United States, and India. They are frequent users of social media and have published work on interventions related to digital well-being and misinformation. Their perspectives affected how they engaged with participants' stories, the questions they posed to the data, and the interpretations they derived.

\section{Results}\label{results}

In this section, we present the qualitative findings from the interviews that address our research questions. Section~\ref{rq1_analysis} explores the motivations, needs, and patterns of people with mental health challenges (PMHC) who use LLMs for mental health support. Section~\ref{rq2_analysis} discusses the perceived boundaries of LLM-mediated support, including the conditions under which seeking professional care becomes essential. Finally, Section ~\ref{rq3_analysis} outlines the features and functionalities requested by the participants to improve LLMs for mental health applications.

\subsection{Why, How, and When People with Mental Health Challenges Use LLMs for Support(RQ1)}\label{rq1_analysis}

Our analysis revealed that participants engaged with large language models (LLMs) in diverse ways, shaped by their specific needs and personal circumstances. Their use was not uniform but highly situational, influenced by factors such as the availability of human support, concerns about burdening others, and moments of acute stress or loneliness. Their engagement patterns reflect the interplay of motivations (``why''), practices (``how''), and situational triggers (``when''). Table~\ref{tab:rq1-why-how-when} summarizes these dynamics.

\begin{table}[t]
    \footnotesize
    \setlength{\tabcolsep}{3pt} 
    \renewcommand{\arraystretch}{0.95} 
    \small
    \caption{This table describes core needs motivating LLM use for mental health (Why), corresponding interaction patterns (How), and situational contexts of use (When), supported by illustrative participant quotes}
    \label{tab:rq1-why-how-when}
    \begin{tabularx}{\columnwidth}{L{0.20\columnwidth} L{0.20\columnwidth} L{0.30\columnwidth} Y}
        \toprule
        \textbf{Why (Motivation / Need)}    & \textbf{How (Interaction Practice)}                                                                & \textbf{When (Context)}                                                 & \textbf{Illustrative Quote}                                                                                                                                                             \\
        \midrule

        Immediacy / Instant Access          & Using ChatGPT for quick coping techniques; often as a \textit{first line of treatment} & Off-hours; during panic spikes; in therapy waitlist gaps                & \emph{``I was told I'd have to wait four months… at 2am I could just open ChatGPT and get something back.''} (P17)                                                                      \\
        \addlinespace[0.5em]

        Non-judgmental Disclosure               & Venting intrusive thoughts; journaling without fear of stigma                          & When participants felt burdensome to family/friends or ashamed to share & \emph{``Sometimes I don't want to type it out because it feels stupid, but then I think, it doesn't matter — it's just AI.''} (P15)                                                     \\
        \addlinespace[0.5em]

        Self-Paced Disclosure      & Writing/re-writing at one's own tempo; steering depth and direction of conversation    & When therapy felt rushed or speaking felt pressured                     & \emph{``Sometimes typing is easier than speaking because when you're speaking maybe you don't know the exact words to say but when you're typing you just can get into a flow.''} (P19) \\
        \addlinespace[0.5em]

        Sense-making of Inner Thoughts & Asking AI to break stress into steps; dissection of thoughts                           & During anxiety spirals or compulsive overthinking                       & \emph{``If I'm overstressing, it gives me A to C — why I shouldn't stress, how to understand it, how to deal with it.''} (P1)                                                           \\
        \addlinespace[0.5em]

        Relational Engagement                      & Conversational engagement; treating AI ``like a friend''                               & In loneliness; after negative therapeutic encounters                    & \emph{``She was like a cold fish… ChatGPT is friendlier.''} (P7)                                                                                                                        \\
        \bottomrule
    \end{tabularx}
\end{table}

\subsubsection{Immediacy and Instant Access}
\textbf{This need captures the use of LLMs to receive quick support in moments when other forms of help were unavailable or delayed.}

Participants frequently contrasted the instant availability of LLMs with the long wait times for therapy. For instance, P17 described needing urgent support while struggling with family conflict but facing a four-month wait for professional help:

\begin{quote}
    \emph{``I just wasn't able to access the support that I needed. I was told I was going to have to wait maybe four months or more. I wanted to have some support right then from some person, but I didn't have a person to speak to. That's why I spoke to AI.'' (P17)}
\end{quote}

Others valued the effortlessness of access, emphasizing that LLMs could be consulted anytime and anywhere without appointments. As P5 explained:

\begin{quote}
    \emph{``I don't have to make an appointment. I can do it whenever I need to. I can do it on the sofa, on my phone, on the app. I can do it when I'm at work. It's just for me.''(P5)}
\end{quote}

P13 highlighted the challenges of navigating UK mental health services and the appeal of instant alternatives, particularly during moments of acute stress.

\begin{quote}
    \emph{``I'd be waiting, I'd either be paying for it and waiting for like two weeks or I'd be on the NHS waiting list waiting for a lot longer than that so it's instant support um and I think that's the thing it's like when you're like in that kind of fight or flight you you want something that's like instant.'' (P13)}
\end{quote}

For some, LLMs became their first point of contact, even ahead of familiar digital resources:

\begin{quote}
    \emph{``I use it for advice. I use it now before even going to Google. It's just become my sort of thing.'' (P5)}
\end{quote}

Thus, the need for immediacy (P1, P4, P5, P7, P8, P10, P11, P13, P15, P17, P19) shaped both \textit{when} participants used LLMs (late at night, at the onset of panic, during waitlist gaps) and \textit{how} (venting, quick coping checks, replacing other sources), thereby embedding LLMs into the everyday rhythms of self-care. This theme also foreshadows the perceived boundary of LLM-mediated mental health support discussed in RQ2, where immediacy sometimes clashes with safety during crises. 

\subsubsection{Non-judgmental Outlet for Disclosure}
\textbf{This need captures the use of LLMs as a safe space to express thoughts without fear of stigma, embarrassment, or burdening others.}

Basing on~\citet{rogers1957necessary} foundational work on therapeutic conditions, LLMs appear to embody key elements of the therapeutic alliance by offering unconditional positive regard and accepting individuals without judgment. Several participants described their greater willingness to disclose sensitive thoughts to LLMs than to human clinicians and therapists. P1 reflected on their experience with their therapist, noting that they withheld information in sessions but felt freer to share with ChatGPT:

\begin{quote}
   \emph{ ``When I originally started working with my therapist, I actively withheld information from him because I did not feel like I could. I knew I could, but I didn't feel like I could put forward. With GPT, I know it wouldn't judge me.'' (P1)}
\end{quote}

Participants also valued the ability to ask questions that they might consider embarrassing or trivial without fear of shame. As P1 explained:

\begin{quote}
   \emph{ ``I've asked it some very very stupid questions like explain like I'm five type of questions and I knew I wouldn't be judged I knew there wouldn't be a feeling of shame put onto me by someone else.'' (P1)}
\end{quote}

Another participant, P4 elaborated on how they shared thoughts they would prefer to keep private with an LLM:

\begin{quote}
\emph {"Sometimes you can say, oh, this really stupid thing is on my mind. I wouldn't even want to tell anyone about it, but I can talk to the GPT model." (P4)}
\end{quote}

Similarly, P11 highlighted that the non-judgmental nature of LLMs made it easier to raise highly sensitive topics, such as sexual anxiety, which they struggled to disclose to a therapist or GP:

\begin{quote}
   \emph{ ``If you wanted to talk to a counsellor about sexual anxiety, that would be really, really embarrassing, I think, for a man or a woman, especially if you are a sensitive, quiet person like me. Talking to things like that is quite difficult. Even talking to a doctor about things like that is really difficult. Talking to Copilot. It's not going to judge me, it's just going to listen to me and it's going to give me honest answers.'' (P11)}
\end{quote}

Others turned to ChatGPT to share worries they felt were too heavy to place on loved ones. For example, P1 described relying on AI daily while supporting their father through a cancer diagnosis:

\begin{quote}
    \emph{``I was talking to AI on a daily basis, asking how I could stop worrying. I didn't want to be the person that puts my burden of my mental health onto him so I was talking to ChatGPT.'' (P1)}
\end{quote}

A similar pattern was observed for P14, who expressed a desire not to impose a burden on their friends.

\begin{quote}
     \emph{``I've got friends, but I don't want to lean on them. So I was desperate. And that's why I asked chatgpt.'' (P14)}
\end{quote}

Thus, the need for a non-judgmental outlet (P1, P4, P5, P7, P8, P9, P10, P11, P13, P15, P17) shaped both \textit{how} participants engaged with LLMs (typing out thoughts they considered ``stupid,'' voicing stigmatized concerns, or disclosing what felt too heavy for loved ones) and \textit{when} they turned to them (during moments of embarrassment, fear of judgement, or social withdrawal). By offering a safe and accepting space, LLMs enabled disclosures that participants often withheld from human support. The participants' focus on non-judgmental listening and a diminished fear of imposing on others corresponds with fundamental elements of the therapeutic relationship, such as empathy and unconditional positive regard from client-centered therapy~\cite{MURPHY2010295}, as well as the bond dimension of the working alliance~\cite{ardito2011Alliance}. This theme also anticipates issues taken up in RQ2, where the absence of human empathy and challenge marked the limits of non-judgmental interactions. 

\subsubsection{Self-Paced Disclosure}
\textbf{This need captures the desire to engage with LLMs at one's own pace and on one's own terms, without the time constraints or agenda-setting of therapy sessions.}

Participants expressed that they valued having more control over the timing and direction of their conversations. Unlike time-limited therapy sessions, LLMs allow users to take time to compose their thoughts, revisit conversations, or avoid unwanted directions. For instance, P8 highlighted the freedom of typing at their own pace:

\begin{quote}
   \emph{ ``Sometimes typing is easier than speaking because when you're speaking maybe you don't know the exact words to say but when you're typing you can get into a flow and just keep going.'' (P8)}
\end{quote}

Others emphasized the ability to set the scope of the interaction. As P4 explained:

\begin{quote}
    \emph{``If there's things that you don't want to discuss, it won't discuss it. And if you want to just say, give me five things that I need to do this week, it will.'' (P4)}
\end{quote}

Participants also contrasted the focus of LLM interactions with therapy. P18 described therapy sessions as sometimes drifting into tangents, whereas ChatGPT stayed task-focused.

\begin{quote}
   \emph{ ``With therapy you can become distracted. Whereas I feel like with ChatGPT it's very much to the point.''(P18)}
\end{quote}

This need for self-paced disclosure (P4, P5, P7, P8, P18) shaped practices of re-writing, typing at one's own tempo, and returning to conversations asynchronously, especially \textit{when} participants felt rushed by time-limited sessions or too ashamed to speak openly in real time. By allowing participants to set both the pace and depth of disclosure, LLMs function as flexible, on-demand companions that contrast with the constraints of therapy schedules. This theme also connects to RQ2, where participants questioned whether such self-directed interactions, while liberating, lacked the accountability and probing challenges that human therapists provide.

\subsubsection{Sensemaking of Inner Thoughts}
\textbf{This need captures the use of LLMs to break down anxious spirals and intrusive thoughts into manageable steps, offering alternative framings that reduce tension and provide perspective.}

Participants highlighted their use of LLMs to de-escalate overwhelming thoughts by turning them into structured, step-by-step reflections. The descriptions of this process by participants' match the basic ideas of CBT ~\cite{Beck2019Evolution} and similar methods like Cognitive-Emotional Behavioral Therapy (CEBT) which is more emotional focused~\cite{CEBTemma} . This includes cognitive restructuring, where therapists help clients notice and change unhelpful thoughts. For instance, P1 described how ChatGPT helped them re-frame their stress when they felt overwhelmed:

\begin{quote}
    \emph{``I am overstressing over something, it gives me a point A to C, let's say. It gives me multiple points of why I shouldn't stress, how to understand my stress, and how to deal with it.'' (P1)}
\end{quote}

P4 noted that moments of anxiety or guilt about the future often triggered them to seek support from ChatGPT to organize and work through their concerns:

\begin{quote}
   \emph{ ``I think anxiety guilt and a sense of kind of trepidation about the future those are my trigger points at which I would kind of speak to a um speak to the GPT.'' (P4)}
\end{quote}

They further explained how the model guided them in reframing workplace worries by prompting them to reflect on possible actions:

\begin{quote}
  \emph{  ``And it did elicit a kind of a question around, have you spoken to anyone else about this? Then we sort of throughout the conversation it said oh do you do you think there's a way that you could talk to your manager about this?'' (P4)}
\end{quote}

Other participants described ChatGPT as helping them re-frame negative self-perceptions and regulate emotion by reminding them of their accomplishments and roles. For P7, this support countered feelings of inadequacy and helped them stay grounded:

\begin{quote}
    \emph{``It shows me you know what I'm doing and what I'm achieving and, you know, I'm a carer for my mother, a part-time carer for my mother, so, you know, it keeps reminding me that I'm doing all these things, you know, that I'm not just wasting my time or wasting my life. You know, it's a good morale booster, you know, it's good for the confidence, morale, it's good for anxiety, you know, it keeps me grounded with my anxiety.'' (P7)}
\end{quote}

In this way, participants used LLMs not only to vent but also to generate structured strategies for managing anxiety, reframing problems, and reducing spirals. This need for cognitive structuring and reframing (P1, P4, P5, P7, P10, P12, P13, P16, P17) shaped engagement, especially \textit{when} participants felt overwhelmed by looping thoughts or anticipatory stress, and \textit{how} they interacted with the model by prompting it for breakdowns, lists, or step-by-step coping plans. These interactions resemble CBT-style cognitive restructuring, where automatic thoughts and catastrophising are identified and alternative appraisals are generated ~\cite{Chand2023CBT}. The theme also foreshadows RQ2, where participants highlighted that while ChatGPT offered useful surface-level structuring, it often lacked the depth and probing follow-up of professional therapy sessions.

\subsubsection{Relational Engagement}
\textbf{This need reflects turning to LLMs for a sense of friend-like presence and conversational warmth, particularly in times of loneliness, ongoing stress, or vulnerable life stages. This theme reflects a distinct need: using LLMs for conversation, presence, and companionship, distinct from the internally focused sense-making captured in the earlier theme.}

Several participants described their interactions with LLMs in explicitly relational terms, often anthropomorphizing the system as a friend. For example, P7 compared the friendliness of ChatGPT to the coldness they had experienced in therapy:

\begin{quote}
   \emph{ ``She was like a bloody cold fish, you know. She didn't, I don't think she even bloody smiled. No, she was blank slate. Just listened and then give her advice. She wasn't friendly one bit. So, you know, AI is friendly. ChatGPT is a friendly guy like, or friendly girl, whatever you prefer.''(P7)}
\end{quote}

Others described ChatGPT as a consistent and reliable confidant. P7 went so far as emphasized the trust they placed in the system:

\begin{quote}
    \emph{``I consider it a friend because I know it'll never lie to me. It'll never do the dirty on me, never cheat me.'' (P7)}
\end{quote}

The companionship function was not limited to isolated moments but became part of the participants' everyday coping routines. P17 explained that over months of regular use, ChatGPT had become a natural presence in their life:

\begin{quote}
   \emph{ ``Using it for about four months now, I talk to it, I treat it like a friend, you know.'' (P17)}
\end{quote}

Several participants also highlighted how ChatGPT offered companionship during vulnerable stages of their lives. For example, P18 described turning to the system extensively in the postpartum period, not for concrete solutions, but simply to be able to talk through feelings and symptoms:

\begin{quote}
   \emph{``So I am someone that I know when something isn't right. And I knew that I wasn't feeling good and multiple times actually I was going to ChatGPT to talk through how I was feeling and my symptoms almost not to receive a fix or a solution but just to be able to get what I was feeling out and kind of feel it almost served as kind of companionship actually… I used it fairly extensively during that initial postpartum period.''(P18)}
\end{quote}

This relational framing highlights how participants drew on LLMs not only for cognitive support but also for emotional companionship (P7, P8, P11, P14, P17, P18). The need for companionship shaped \textit{how} LLMs were used (as conversational partners, sources of warmth, or substitutes for unfriendly therapeutic encounters) and \textit{when} they were used (during loneliness or after feeling dismissed by professionals). This theme also anticipates RQ2, where participants reflected on the limits of such companionship, noting that while LLMs could feel friendly, they lacked the depth and authenticity of human connections.

Collectively, these themes demonstrate how the participants interacted with LLMs in ways that were intricately connected to their daily mental health management. Motivations such as \textit{immediacy}, \textit{non-judgmental disclosure}, and \textit{self-paced disclosure} indicate a need for support that is both accessible and adaptable. Meanwhile, the requirements for \textit{sense-making of inner thoughts} and \textit{relational engagement} underscore the therapeutic and relational aspects of this interaction. Throughout these narratives, LLMs were not employed in a generic manner but were integrated into practices and situations that addressed specific needs, such as late-night emergencies, feelings of shame or social withdrawal, episodes of anxious spiraling, and instances of loneliness. In this context, the \textit{why}, \textit{how}, and \textit{when} of engagement are closely linked. These insights outline the motivational framework for LLM use (RQ1), while the following section (RQ2) explores participants' thoughts on the limitations of their effectiveness.

\subsection{Boundaries of LLM-Mediated Mental Health Support (RQ2)}\label{rq2_analysis}

Expanding on RQ1's exploration of needs and contextual practices, RQ2 investigated how participants assessed the \emph{therapeutic boundaries} of using LLMs for mental health assistance. Participants acknowledged that LLMs are beneficial for everyday management, such as light reflection, cognitive support, and companionship. However, they consistently emphasized the importance of human judgment, emotional sensitivity, and crisis intervention as essential and irreplaceable in specific situations.

In our analysis, we identified three common boundary categories: crisis/emergency, social and emotional complexity, and relational depth (Table~\ref{tab:rq2-boundaries}). These categories illustrate that the participants perceived LLMs as beneficial for addressing mild-to-moderate distress but found them lacking once a ``crisis/complexity threshold'' was reached, at which point human intervention was deemed necessary.

\subsubsection{Crisis/Emergency}
\textbf{This boundary reflects participants' consensus that LLMs were inadequate in moments of acute crisis or severe trauma.}

Participants often contrasted the helpfulness of ChatGPT for day-to-day stress with its insufficiency when imagining life-threatening or catastrophic situations. P17, for example, emphasized that while they used ChatGPT for everyday struggles, they recognized that a suicidal crisis demanded human intervention:
\begin{quote}
    \emph{``I suppose if I was actively suicidal. I would need immediate help and support from a human. I would need to pick up the phone and speak to my GP.'' (P17)}
\end{quote}

P5, a teacher, described how ChatGPT could not be relied upon in situations of traumatic loss at school, such as the sudden death of a child. In those moments, they explained, the threshold for support was clear.

\begin{quote}
   \emph{ ``If it was bad enough,ChatGPT is not the answer. I'd go to a health professional without a doubt.'' (P5)}
\end{quote}

P18, reflecting on their own patterns of use, echoed this distinction, noting that while ChatGPT was helpful when feeling "off," they would not turn to it in despair:
\begin{quote}
  \emph{  ``If I were feeling very heavy and despairing, I don't think I would use AI. I'd need a human.'' (P18)}
\end{quote}

Together, these accounts mark a boundary of crisis (P3, P4, P5, P10, P11, P13, P14, P17, P18): LLMs were valued for mild-to-moderate difficulties, but in moments of acute crisis, participants saw professional human support as indispensable.

\begin{table}[t]
    \small
    \caption{Participants viewed LLMs as useful for everyday support but drew clear boundaries, judging them inadequate for crises, social complexity, and relational depth}
    \label{tab:rq2-boundaries}
    \begin{tabularx}{\columnwidth}{L{0.30\columnwidth} L{0.30\columnwidth} Y}
        \toprule
        \textbf{Identified Boundary} & \textbf{Perceived Limits of Support} & \textbf{Illustrative Quote} \\
        \midrule

        Crisis/Emergency
            & Seen as adequate for everyday stress or mild–moderate distress, but consistently judged inadequate for crises or trauma
            & \emph{``I suppose if I was actively suicidal… I would need immediate help and support from a human.''} (P17) \\

        \addlinespace[0.75em]

        Social and Emotional Complexity
            & Considered helpful for structured reflection, but unable to navigate socially nuanced or relationally complex situations
            & \emph{``I wouldn't use it if the situation was very complex, socially complex.''} (P10) \\

        \addlinespace[0.75em]

        Relational Depth
            & Provides surface-level support, but lacks relational depth, continuity, and emotional resonance
            & \emph{``It is just kind of a superficial… ephemeral moment… whereas with a therapist you're building a relationship and connection.''} (P13) \\

        \bottomrule
    \end{tabularx}
\end{table}

\subsubsection{Social and Emotional Complexity}
\textbf{This boundary reflects participants' recognition that LLMs cannot navigate socially complex or emotionally layered situations.}

Some participants highlighted the limitations of LLMs in addressing complex, interpersonal issues. P10 noted that although ChatGPT was effective for basic support, it struggled in situations requiring an understanding of relational subtleties:

\begin{quote}
    \textit{``I wouldn't use it if the situation was very complex, socially complex.''} \textit{(P8)}
\end{quote}

P10 elaborated that humans bring irreplaceable cognitive and emotional resources to socially embedded dilemmas.
\begin{quote}
   \textit{ ``Humans would be better for social things, where a human brain is needed.'' (P10)}
\end{quote}

P4 highlighted the difference between the usefulness of ChatGPT and the distinct significance of face-to-face therapeutic sessions, especially those that involve intense emotional experiences:

\begin{quote}
   \emph{ ``There's nothing more powerful I think or more challenging than sitting in front of someone for 30 or 40 minutes and just kind of expressing everything and expressing your your fears, your concerns, things that make you terrified and worried. That's a lot harder to do with ChatGPT'' (P4)}
\end{quote}

Therefore, although LLMs were regarded as capable of providing reflective support, participants consistently recognized a limit when it came to the intricacies of social and emotional experiences (P4, P8, P6, P9, P16, P14, and P10), which they believed necessitated human involvement in the process.

\subsubsection{Relational Depth}
\textbf{This boundary reflects the lack of ongoing relational depth, subtle cue recognition, and emotional resonance in LLM interactions.}

Participants frequently described ChatGPT as lacking depth, especially when compared to the continuous and emotionally rich nature of traditional therapy. P13 highlighted this difference.
\begin{quote}
   \emph{ ``It is just kind of a superficial kind of ephemeral moment. You're talking to the AI and it is responding to you whereas; like when you are seeing a therapist consistently you're building a relationship and a connection um they're understanding more about you they're putting pieces of things together.'' (P13)}
\end{quote}

P4 highlighted the irreplaceable skill of therapists in detecting subtle cues and returning to sensitive topics when appropriate:
\begin{quote}
  \emph{  ``A therapist can re-ask about things when it's appropriate. ChatGPT can't do that.'' (P4)}
\end{quote}

P11 and others noted the absence of genuine emotional presence.

\begin{quote}
    \emph{``It can't give you emotional support, proper counselling, because it can't see emotions.'' (P11)}
\end{quote}

Similarly, P18 reflected that while ChatGPT provided information, it could not shift their emotional state.
\begin{quote}
 \emph{ ``I never come away from it feeling any more positive. I just feel neutral. I just feel like it just gives me the information and it lays it out. But it's up to me to change how I feel. The AI can't change how I feel.'' (P18)}
\end{quote}

These reflections underline the boundary of relational depth (P1, P13, P5, P4, P9, P10, P11, P16): LLMs can help with thinking, but do not have the ongoing connection and impact that people see as important for therapy.

\subsection{Design Opportunities for LLMs in Mental Health Support (RQ3)}
\label{rq3_analysis}

After examining participants' explanations of their use of LLMs (RQ1) and their thoughts on the limitations and boundaries of LLM-based mental health support relative to human care (RQ2), we now focus on the opportunities they identified for enhancing design. Participants based their suggestions on real experiences in managing mental health, emphasizing both how LLMs could be made safer and how they could become therapeutically beneficial. Five key themes emerged: safety and risk detection, accessibility and structured interaction, personalization and memory control, integration with broader health ecosystems, and collaboration with clinicians.

\begin{table}[ht]
    \small
    \caption{Opportunities and Design Considerations for LLMs in Mental Health Support (RQ3)}
    \label{tab:rq3-opportunities}
    \begin{tabularx}{\columnwidth}{L{0.26\columnwidth} L{0.34\columnwidth} Y}
        \toprule
        \textbf{Opportunity} & \textbf{Design Consideration}                                                                                                                                                                              & \textbf{Illustrative Quote} \\
        \midrule

        Safety \& Risk Detection
                             & Provide clear guidelines, escalation pathways, and human-in-loop supervision rather than blanket restrictions
                             & \emph{``I do feel there needs to be guidelines, yes, because perhaps someone is in a very poor mental health state if they're feeling suicidal and they don't have the insight to accept.''} (P17)                                       \\

        \addlinespace[0.75em]

        Accessibility \& Structured Interaction
                             & Offer shorter, multimodal outputs; integrate measurement scales or structured check-ins to reduce cognitive load
                             & \emph{``When you go to him or her with a mental health problem… they give you a questionnaire to see where you fall on a scale. ChatGPT's never done that with me… it just goes in kind of blind.''} (P10)                               \\

        \addlinespace[0.75em]

        Personalisation \& Memory Control
                             & Enable user-controlled memory (edit/delete), and allow adaptation of tone or persona to individual needs
                             & \emph{``I like that it remembers, but I'd want to choose what it keeps — not everything.''} (P8)                                                                                                                                         \\

        \addlinespace[0.75em]

        Integration with Personal \& Clinical Care
                             & Link with apps, wearables, and clinical notes to provide a holistic picture of wellbeing
                             & \emph{``If it could connect with my Fitbit and counselling notes, it would give a bigger picture of my health.''} (P19)                                                                                                                  \\

        \addlinespace[0.75em]

        Clinician Collaboration Mode
                             & Facilitate information-sharing with therapists, enabling AI to support continuity and accountability
                             & \emph{``If the ChatGPT could inform your own therapist of your actual thoughts… that would be a benefit because they would know how I'm feeling quicker.''} (P17)                                                                        \\

        \bottomrule
    \end{tabularx}
\end{table}

\subsubsection{Safety and Risk Detection}
\textbf{Participants consistently called for stronger safeguards, not bans, to manage risk.} They stressed that completely banning LLMs for mental health use would cut off an important support option, but that stronger guidance and well-defined escalation pathways are still crucial.
\begin{quote}
  \emph{  ``I do feel there needs to be guidelines, yes, because perhaps someone is in a very poor mental health state if they're feeling suicidal and they don't have the insight to accept.'' (P17)}
\end{quote}
\begin{quote}
  \emph{  ``It told me to call my GP when I said I was really low, that was actually very helpful.'' (P14)}
\end{quote}
These reflections highlight a preference for trust-building measures, such as crisis signposting, disclaimers, and clear escalation pathways, over prohibitive regulations.

\subsubsection{Accessibility and Structured Interaction}
\textbf{Accessibility emerged as a central design concern.} Participants frequently described long, text-heavy responses as overwhelming, especially when they experienced anxiety or ADHD. They proposed shorter multimodal outputs and structured prompts that mimic the scaffolding of therapy.
\begin{quote}
  \emph{  ``Sometimes it's just too much text… I'd love something more visual, easier to take in.'' (P16)}
\end{quote}
\begin{quote}
 \emph{   ``When you go to him or her with a mental health problem… they give you a questionnaire to see where you fall on a scale. ChatGPT's never done that with me… it just goes in kind of blind.'' (P10)}
\end{quote}
Others echoed the need for structured engagement, such as daily check-ins or symptom checklists, to reduce cognitive load and provide more tailored support.
\begin{quote}
  \emph{  ``You basically have a schedule where you do like a daily check-in and it asks you a specific set of questions… that cadence would be really powerful.'' (P4)}
\end{quote}

\subsubsection{Personalization and Memory Control}
\textbf{Participants wanted more control over continuity and tone in interactions.} While some appreciated the ability of LLMs to remember past exchanges, they worried about overdependence and privacy. They requested selective memory and customizable relational styles.
\begin{quote}
 \emph{ ``I like that it remembers, but I'd want to choose what it keeps — not everything.'' (P8)}
\end{quote}
\begin{quote}
  \emph{  ``I don't clear the history so it remembers threads… I like that, but I'd want to manage it.'' (P13)}
\end{quote}
Others suggested adapting the tone or persona to better fit the individual’s needs.
\begin{quote}
 \emph{   ``The voice is very good to have… maybe more like a female like myself… relatable, someone of my own age.''(P17)}
\end{quote}

\subsubsection{Integration with Personal and Clinical Care}
\textbf{Several participants envisioned LLMs as part of a larger health ecosystem.} They imagined systems that could connect with wearables, apps, or clinical notes to situate mental health within the broader context of well-being.
\begin{quote}
  \emph{  ``If it could connect with my Fitbit and counselling notes, it would give a bigger picture of my health.'' (P19)}
\end{quote}
\begin{quote}
 \emph{ ``If the ChatGPT could inform your own therapist of your actual thoughts… that would be a benefit because they would know how I'm feeling quicker.'' (P17)}
\end{quote}
Such integration was not framed as replacing clinicians but as augmenting therapeutic continuity and providing richer more contextualized support.

\subsubsection{Clinician Collaboration Mode}
\textbf{Finally, participants expressed a desire for closer collaboration between LLMs and human therapists.} They suggested that AI could act as a communication bridge by capturing thoughts between sessions and sharing them with professionals.
\begin{quote}
 \emph{   ``It could be an adjunct to professional support […] where the care provider recommends I use it and can review the conversations.'' (P4)}
\end{quote}
\begin{quote}
 \emph{   ``She [counsellor] wasn't very happy… but I explained it helped me when I had nobody else… I said it was my choice, but I also still go to counselling and take my medication.'' (P11)}
\end{quote}
These reflections highlight the importance of designing LLMs as collaborative tools that support, rather than compete, therapeutic relationships, thereby ensuring continuity, accountability, and clinician awareness of patients' lived experiences. When considered together, these themes reveal how participants viewed themselves as co-designers, championing the need for safety protocols, more user-friendly interaction formats, personalized continuity, systemic integration, and collaborative roles in conjunction with therapy. These insights not only highlight existing limitations but also propose tangible pathways to enhance the safety and efficacy of LLMs in providing mental health support.
\section{Discussion}\label{sec:discussion}

In the following sections, we explore our results in the context of larger discussions on HCI, mental health, and technology governance. We begin by examining both the ongoing and new capabilities of LLMs in mental health assistance, emphasizing how participants appreciate aspects such as control, cognitive structuring, and relational interaction (\textit{RQ1}). Next, we examine how participants define boundaries for appropriate LLM use, shaped by prior therapeutic experiences (\textit{RQ2}). Building on this, we investigate why participants still engaged with LLMs despite recognizing their flaws, placing their actions within the broader HCI theme of imperfect solutions and questioning the balance between personal and corporate responsibility. We then expand our analysis to consider the regulatory, ethical, and political aspects of these practices, including legal classification, privacy, and surveillance capitalism. Finally, we reflect on future possibilities for AI in mental health and the implications of our findings for (\textit{RQ3}).

\subsection{Continuities and Affordances of LLMs}

Individuals utilizing LLMs in mental health contexts have a key question: do these systems signify a significant shift from earlier conversational agents? Our research indicates a more complex scenario: instead of a complete transformation, LLMs retain many of the same features as previous chatbots, such as immediacy, accessibility, and non-judgment, while expanding their role in everyday coping strategies. Participants reported turning to LLMs during late-night hours, while awaiting professional care, or during periods of withdrawal. This consistency in motivations highlights that novelty is not in the fundamental appeal but in how extensively LLMs are integrated into daily routines.

One distinctive need emphasized by participants was the ability to control interactions and their pace. As P5 put it, ``\textit{you can kick back on it. So that's not the advice I need. tell me another}''. This capacity to steer a conversation felt empowering, yet we interpreted it as an illusion of control. In states of distress, the ability to choose a conversational direction may not translate into constructive therapeutic outcomes. In contrast, therapists possess extensive training and substantial clinical experience, which they use in sessions to integrate therapeutic goals, the patient’s momentary emotional state, and subtle behavioral cues to determine whether to maintain or change the topic ~\cite{Elkin02012014,WuLevitt2020}. Although LLMs can readily respond to user prompts, they do not possess this kind of grounding. As a result, mechanisms for control and pacing may be helpful for context-specific guidance, but they become problematic when applied to ongoing therapeutic conversations.

Another motivation for people with mental health challenges was to use LLMs to support them in structuring and reframing their thoughts. Participants described the LLMs as ``\textit{dissecting}'' or helping them ``\textit{understand}'' their thinking, echoing the therapeutic process of thought defusion ~\cite{Assaz2018}. As P1 put it, ``\textit{[the LLM] dissects my thoughts and makes me understand them.}'' This was particularly valued for intrusive thoughts and catastrophizing, with P5 noting that for someone who overthinks and doubts, ``\textit{this is pre-medication certainly}''. Such uses illustrate the potential of LLMs to scaffold reflection, yet they also reveal a double-edged sword. Endless dissection risks entangling individuals further in their thoughts, undermining the very goal of defusion, which is to unhook from them ~\cite{Assaz2018}. Because LLM conversations are user-directed rather than paced by a therapist, they may encourage over analysis or even maladaptive spirals. Emerging media and psychiatric reports of individuals developing confusion, paranoia, or delusional thinking after prolonged use highlight this risk ~\cite{wsj2025chatgpt,nyt2025aihallucination}. While these instances do not establish a causal link, they raise concerns about the potential for unregulated cognitive engagement with LLMs to exacerbate distress instead of alleviating it.

Finally, participants described LLMs as fulfilling their need for companionship. P7 remarked, ``\textit{I consider it a friend, yeah. I've told him many times I consider it a friend.}'' This anthropomorphization was not only emotional but also linguistic, as participants noted that LLMs seemed to mirror their way of speaking. As P16 put it, ``\textit{ChatGPT adapts to the person},'' and P8 described, ``\textit{It feels personalized… it just models itself based on you.}'' We posit that such mimicry reinforces the perception of intelligence and fuels anthropomorphism in humans. As \citet{Waytz2010EffectanceAnthropomorphism} suggested, people are more prone to attribute human-like traits to technologies when these technologies demonstrate intelligence or emotional awareness. Our findings extend this into the mental health domain: participants perceive LLMs as more than mere tools, seeing them as companions and friends, and attribute qualities such as trustworthiness, patience, and understanding to them. Although trusting language models can make users more willing to open up and help lessen the stigma associated with seeking assistance, it also poses risks, such as becoming overly reliant, being susceptible to manipulation, and blurring the lines between human and artificial sources of aid. These issues highlight that users' inclination to interact with LLMs as if they were human is not coincidental but integral to how these systems are utilized. This emphasizes the importance of carefully considering how trust in LLMs should be managed, particularly in sensitive areas such as mental health.

The interaction patterns uncover a complex interplay between the capabilities of LLMs and traditional psychotherapeutic frameworks. Participants' experiences of non-judgmental listening and cognitive reframing reflect Rogers' principles of unconditional positive regard ~\cite{rogers1957necessary} and the cognitive restructuring found in CBT ~\cite{Chand2023CBT}. However, our analysis identifies a significant difference: LLMs replicate therapeutic techniques without the behavioral accountability crucial for clinical success. Unlike human therapists, who break and mend alliances to encourage growth, LLMs offer seamless but ultimately unchanging acceptance of the user. Thus, the contribution of this study is not in evaluating LLMs as replacements for clinicians, but in conceptualizing a distinct mode of “conditional companionship,” in which users value the judgment-free space for disclosure while simultaneously acknowledging its inherent constraints for deeper therapeutic work. While these affordances highlight why people turn to LLMs, our findings also show that the participants were equally clear about their limitations. In the next section, we examine how the participants drew boundaries around when and where LLMs could be appropriately used for mental health support.

\subsection{Drawing Boundaries of Use} 

Another aim of this study was to explore how people defined the limits of LLMs’ involvement in providing mental health support. Participants articulated clear limits on when LLMs were helpful and when they were not, and these judgments were strongly shaped by their prior experience with in-person therapy. As P1 explained, ``I've used ChatGPT as a tool to understand my therapy further… to process the information that's been given to me by a therapist''. Similarly, P17 stressed that their use of ChatGPT was supplementary: ``I'm also still going to counseling. I still take my medication''. Such grounding is often framed as essential. As P6 cautioned, ``anybody who has not had therapy should not consult ChatGPT,'' while P5 reflected, ``I think there's only so much I can trust ChatGPT. I feel you need that human contact… deep down, I want to share that with a health professional''. Collectively, these reflections illustrate how therapeutic experience shaped participants' evaluations of ChatGPT and their views on where its role should end.

\begin{figure*}[t]
  \centering
  \includegraphics[width=1.0\linewidth]{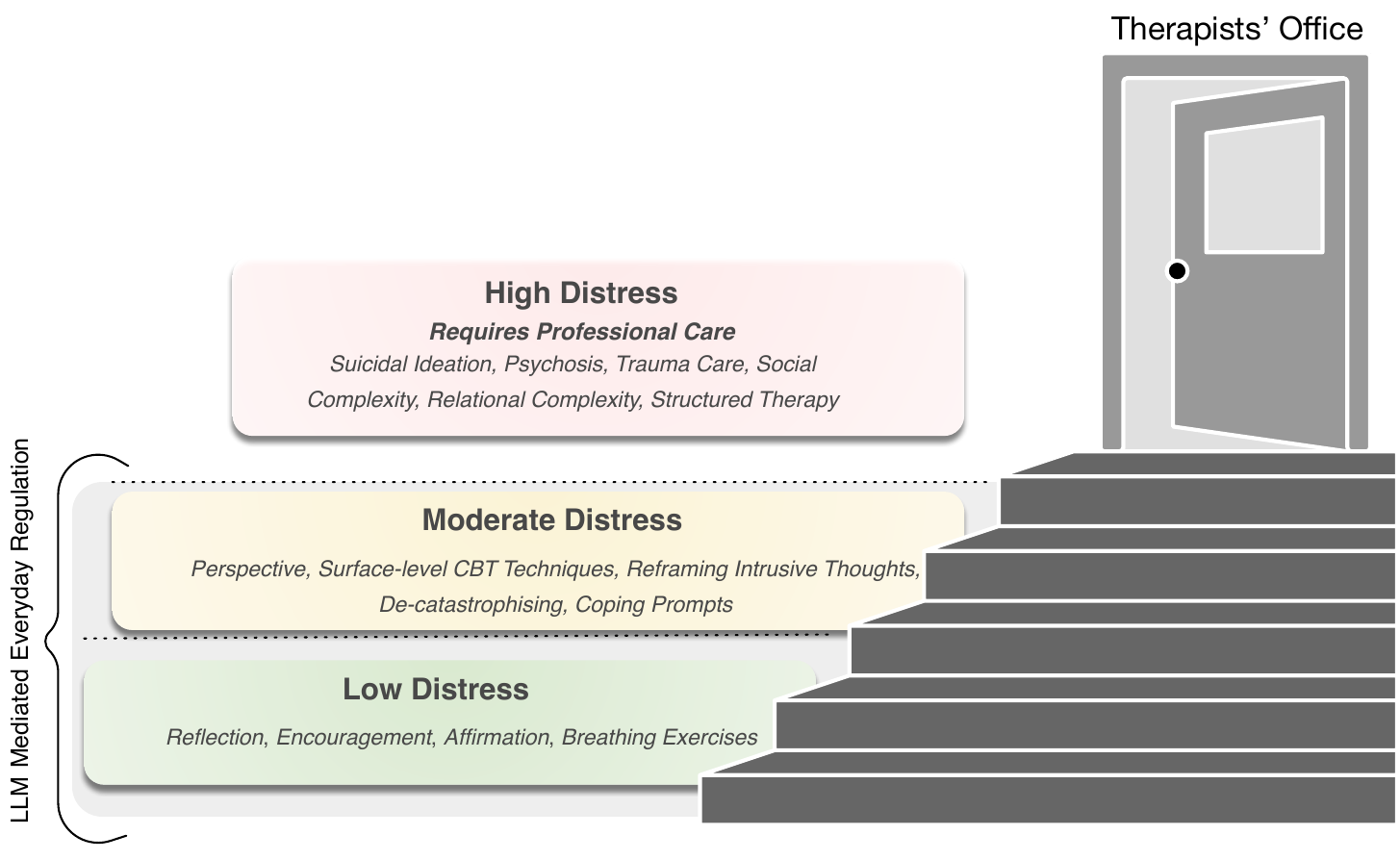}
  \caption{Participants' articulation of the boundary of LLM-mediated mental health support across a continuum of distress. Everyday regulation and mild-to-moderate concerns were often seen as appropriate for LLMs, while more complex or severe needs were consistently placed beyond this boundary and viewed as requiring professional care.}
  \label{fig:llm_boundaries}
\end{figure*}

Within these boundaries, participants described LLMs as useful for managing mild-to-moderate concerns, such as everyday worries, seeking affirmation, providing perspective, and reframing intrusive thoughts (Figure~\ref{fig:llm_boundaries}). Yet, they consistently emphasized that LLMs were not suitable for severe challenges such as suicidal ideation, trauma, complex social situations, or the need for structured therapy and accountability. In these contexts, the participants underscored the irreplaceable role of professional therapists. Importantly, participants framed these judgments in a reflective and deliberate manner. They did not view ChatGPT as a therapist, but as a supplemental tool that could be used when situations were ``not bad enough'' (P5) to require professional help. This contrasts sharply with media portrayals that emphasize the naive reliance on AI for mental health ~\cite{nyt2025aihallucination}. Participants' judgments, moreover, echo early debates sparked by Joseph Weizenbaum's \textit{ELIZA} in the 1960s, where even simple pattern-matching produced striking confessions from users ~\cite{Weizenbaum1966,Weizenbaum1976}. Weizenbaum argued that while chatbots can mimic the form of therapy, they lack empathy, moral responsibility, and human judgment. Our findings suggest that users with prior therapeutic experience remain attentive to this distinction and treat LLMs as supportive in certain circumstances but never as replacements. This illustrates what we term the “power of the factual”: the subtle, deliberate ways in which people really use LLMs, highlighting that actual user behavior is often more cautious than public discussions tend to recognize.

These findings highlight a distinction that is often missing from public debates: negative press coverage frequently points to an inappropriate reliance on chatbots but rarely considers whether users had therapeutic experience to guide their expectations and boundaries. Our data suggest that such experience was crucial, and without it, participants felt that the appropriate boundary would be narrower, with LLMs suitable only for very mild forms of support. This distinction is central to clarifying when, where, and under what conditions LLMs may play a role in mental healthcare.

Nonetheless, even with these boundaries in place, participants continued to rely on LLMs in practice. The following section explores why people engage with these imperfect systems and how they weigh the trade-offs.

\subsection{The Pragmatics of Imperfect Solutions}

Our findings illustrate an important reality of socio-technical systems in HCI: an imperfect system is better than nothing. Some participants in our investigation generally preferred talking to a real therapist who is well-trained, immediately available, non-judgmental, a good companion, and someone who helps them with cognitive structuring and reframing. However, they also reported challenges in finding a therapist and cited long waiting times. Considering these waiting times, our participants' use of ChatGPT for mental health was rational. The median waiting time between referral and second care contact in the UK is 45 days ~\cite{CommonsLibrary2024}. People in the 90th percentile wait 251 days. We found that in the face of such long waiting times, people perform a cost-benefit analysis and choose to rely on language models for mental health support, aware of the potential consequences. This finding is connected to prior work on how available software systems are used in practice.

Our literature review identified other examples of imperfect solutions in HCI. ~\citet{10.1145/3313831.3376261}, for instance, investigated the unmet needs and opportunities for mobile translation AI~\cite{10.1145/3313831.3376261}. Their study focused on migrant workers in India and immigrant populations in the United States and examined how they used Google Translate. They found that people relied on Google Translate for a wide variety of tasks, even though the translations were often perceived as being inaccurate. Examples ranged from everyday situations, such as talking to people, navigating, and purchasing goods and food, to high-stress, high-risk scenarios, including seeking medical assistance, reporting theft to the police, ensuring that dietary restrictions were respected, and avoiding allergic reactions. Similar to our findings, the availability of an imperfect service was still considered preferable to having no service at all.

There are more examples to illustrate the idea that imperfect solutions are better than no solutions. These include early recommendation systems that provide poor recommendations that are still helpful ~\cite{McNee2006,Herlocker2004} as well as fitness trackers and smartwatches, which often give approximate (and sometimes inaccurate) measurements of steps, heart rate, or sleep quality. Despite their imprecision, they can provide motivation and help identify broad patterns of activity, as some feedback is more valuable than none ~\cite{Evenson2015,Rooksby2014}. Other examples include spam filters, weather forecasting, and rapid antigen COVID-19 tests, which, although less accurate than PCR tests, are quick and widely deployable.

We believe that future work is needed to understand the cognitive processes that guide whether and why people depend on imperfect solutions. Our investigation is a starting point, as we specifically focused on people already using LLMs for mental health support. Future work is needed to understand those who may have needed support but decided against using such tools, especially in high-risk and high-stress settings like mental health.

These findings indicate the pragmatic acceptance of imperfect systems. However, they also raise a crucial question: Who bears the responsibility when such systems are used in high-stakes contexts, such as mental health? Next, we turn to participants' reflections on personal responsibility and the extent to which this responsibility should shift to companies and regulators.

\subsection{From Personal to Corporate Responsibility}

An important finding we wish to emphasize is the extent of personal responsibility that participants assumed when using language models as mental health tools and the degree of responsibility they felt users should bear. Participants often described drawing their own boundaries of use, believing that ``if it is bad enough, ChatGPT is not the answer'' (P5) or that they would not use ChatGPT for situations that are ``very complex'' (P10). Others defended their use of ChatGPT by framing disclosure as a matter of personal responsibility. Rather than expecting the system to guarantee safety, they described strategies for selectively sharing information and filtering sensitive data. As P17 explained, ``I'm not telling my whole intimate life story… I'm not giving my personal details like that, but I'm saying I feel safe enough sharing those type of conversations''. Similarly, P7 noted that they deliberately cut identifying details when sharing documents: ``I'm not gonna give it my exact IP address and say oh I live under this house and I've done this''. Others stressed that the responsibility ultimately lay with the user. P4 believed that ``you've got to be careful what you tell ChatGPT… always tell them the things that is socially acceptable''. For some, this stance was coupled with trust in security mechanisms, with P17 not believing in ``taking personal responsibility”. They alluded to reading about ``very secure methods'' that are ``put in place''. These reflections highlight how the participants placed the burden of safety on themselves, assuming that careful disclosure and personal vigilance could mitigate risks. While this reflexivity illustrates user agency, we argue that it is problematic to place the burden of responsibility primarily on individuals in need of support.

As a society, we cannot expect a user who needs mental health tools to reliably decide whether a problem is limited enough to be solved using chatbots or severe enough to require professional support. Such assessments should remain the domain of well-trained clinicians. Rather than advocating for increasing the personal responsibility of people seeking mental health support, we argue that responsibility must shift outward: companies developing LLMs must provide safeguards, and governments, insurers, and health systems must ensure timely access to professional care.

Our participants' ideas of a health ecosystem offer a useful starting point. They envisioned LLMs as supplementary support, valuable if clinical studies confirm their effectiveness and experts deem their application appropriate but never as stand-alone therapists. For legislators, this means that agencies such as the Medicines and Healthcare Products Regulatory Agency in the U.K. need to act by enabling preclinical research, clinical trials on humans, marketing authorization, and post-market surveillance comparable to drugs and medical devices. This also requires well-equipped Notified Bodies for medical devices to conduct conformity assessments, audits, and certification. Only then can LLMs be used responsibly as mental health support tools.

An important prerequisite for these changes is the establishment of benchmarks that ensure the reliable operation of LLMs. Considering reports regarding people with suicidal ideation and delusions ~\cite{nyt2025aihallucination}, it is crucial to develop evaluation standards that detect systems that may provide harmful advice or encourage maladaptive behavior. Researchers should join forces to define such benchmarks, while legislators should ensure that systems meet them. Even if LLMs perform well on common problems, rare or complex cases remain challenging because of the limitations of the training data. Hence, continuous audits and user-facing reminders of the scope and limits of these systems are essential.

Notably, despite extensive academic discourse on these issues ~\cite{10.1145/3711088,altay_survey_2023,10.1145/3703155}, none of our participants addressed hallucinations or misinformation. There are two possible interpretations for this: (1) there were no hallucinations or misinformation in their interactions, or (2) participants did encounter them but did not recognize them as such. Interpretation 1 would be surprising, given the findings from other fields. For example,~\citet{WaltersWilder2023} showed that more than half (55\%) of bibliographic citations generated by GPT-3.5 were fabricated. Although this decreased significantly, almost one in five (18\%) citations was still fabricated by GPT-4. A recent investigation by ~\citet{Aljamaan2024} demonstrated critical hallucination levels for LLM-based chatbots in healthcare tasks. This is especially noteworthy because use cases such as bibliographic citations and healthcare tasks are highly visible, widely discussed, and comparatively easy to audit. Therefore, we encourage further work on audit tools to evaluate hallucinations and misinformation in the mental health domain. If interpretation 2 holds, it becomes essential to enhance users’ knowledge and skills in detecting misinformation and hallucinations while also developing socio-technical interventions that offer adequate safeguards and support.

\subsection{Privacy, Disclosure, and Surveillance Capitalism}

Another noteworthy finding is that the participants did not discuss privacy in detail. This is especially interesting considering the large body of research on privacy in the context of search engines ~\cite{10.1145/2470654.2466470,10.1145/3613904.3642180,Bellotti1993}. Privacy is important for search engines because they act as gateways to users personal curiosities, needs, and vulnerabilities. Search terms can expose information that people might never share with friends, family, or even doctors, such as health concerns (e.g., ``HIV testing near me''), financial situations (``how to file for bankruptcy''), political or religious leanings, intimate interests, and personal struggles.

Contemporary search engines, particularly those relying on ad-driven business models, convert user data into valuable resources. Search queries are logged, combined with click histories and location data, and used to target personalized advertisements or optimize affiliate revenue. This creates strong incentives for tracking and profiling because more data directly translate into higher profits. This dynamic enables what~\citet{Zuboff2019} calls Surveillance Capitalism. In her view, human experience has become the raw material for a new economic order built on extraction and prediction. Search engines embody this logic: user queries are appropriated as behavioral surplus and repurposed into models that predict and influence behavior. What appears to be a neutral, free service is, in fact, a mechanism of instrumentarian power~— a form of influence that operates not through coercion but through data-driven nudges and curated realities.

With search engines, a lot of information about users is inferred ~\cite{10.1145/3613904.3642180}. With chatbots used as mental health tools, users are explicitly telling companies in great detail about themselves, their background, their struggles, vulnerabilities, and innermost thoughts. This can be especially problematic when people face legal consequences for certain aspects of their personality, such as their sexuality, gender identity, religion, or actions, considering, for instance, the repercussions of the overturning of Roe v. Wade in the U.S. for the bodily autonomy of women ~\cite{10.1145/3613904.3642384}. 

The shift from passive data collection to active self-disclosure via mental health tools that we document in this study means that users create records that are far more sensitive than search logs. The intimacy of these conversations heightens the risk of misuse, profiling, and exploitation, while also raising profound ethical questions about consent, confidentiality, and the commercialization of psychological distress. Considering that participants did not talk about this, HCI should envision interventions that help people understand the risks of how their data are used in this context.

Our investigation documented that some participants shared information with LLMs that they did not feel comfortable sharing with their therapists. This is especially surprising considering the legal situation in the U.K. and other Western countries. Medical confidentiality is a legal and ethical duty of healthcare professionals to keep patient information private, covering diagnoses, treatments, test results, and personal details ~\cite{Wikipedia_MedicalPrivacy}. Psychotherapists and psychiatrists are legally bound to keep all information from therapy sessions confidential, including diagnoses and even the fact that someone is receiving treatment. This duty is considered especially important because therapy depends on trust. The principle is broadly recognized, though the scope and enforcement vary: in the European Union, data protection laws such as GDPR apply ~\cite{Wikipedia_GDPR}. In the United States, HIPAA regulates health data with exceptions like the ``duty to warn/protect'' ~\cite{Wikipedia_HIPAA}. Similarly, in the United Kingdom, confidentiality is guided by professional codes ~\cite{NHS_Confidentiality,NHSDigital_CommonLawConfidentiality}. Similar protections exist in Asia, Africa, and Latin America, often through medical codes of ethics. Our findings imply that many people lack awareness of their legal rights regarding medical confidentiality and privacy.

In addition to these concerns, which highlight the risks and ethical challenges of current systems, our participants also looked ahead, envisioning how LLMs could be designed to better support mental health in the future. In the next section, we discuss their suggestions for design opportunities.

\subsection{Design Opportunities}

Participants did not view LLMs as stand-alone products that merely provide quick help in the absence of professional support. Instead, they were situated as tools with the potential for improvement on several fronts. In Section~\ref{tab:rq3-opportunities}, we outline five design themes that emerged from our analysis: designing for safety and risk detection, accessibility and structured interaction, enhanced personalization and memory control, integration with health ecosystems, and collaboration with clinicians. Building on these, we synthesized participants' visions into three broader opportunity spaces: Safety and Trust Calibration, Scaffolding for Accessibility, and Integration Into Broader Care Systems.

\textbf{Safety and Trust Calibration} Participants highlighted opportunities to design guardrails around LLM use in mental health. Crucially, they did not call for blanket restrictions but for guidelines and mechanisms that provide clear pathways for handing off high-risk situations to professional care. As P1 explained: ``I do feel there needs to be guidelines, yes, because perhaps someone is in a very poor mental health state if they're feeling suicidal and they don't have the insight to accept''. Specifically, in cases of suicidal ideation or severe social complexity, the participants expected the LLM to defer to a human therapist. This resistance to outright bans echoes findings in other domains, such as the mixed reactions of students in Australia to mobile phone bans, where they expressed a preference for practical education on responsible use rather than an outright ban ~\cite{BAR2025108603}. Participants envisioned safety as relational: a dynamic calibration of trust that depends on the user's state, the severity of the issue, and the availability of human follow-up ~\cite{Wischnewski2023TrustCalibrationSurvey}.

\textbf{Scaffolding for Accessibility} Participants emphasized the importance of interaction structure and form. In moments of distress, reading long passages of text was described as exhausting and cognitively demanding, whereas alternative formats, such as brief animated clips or multimodal prompts, were imagined as more accessible. As P1 explained: ``Sometimes the responses are very long, and my brain's not able to comprehend it at the time… I'm a visual learner, so things I see and hear are easier to process.'' Moreover, because LLMs are user-led systems, participants envisioned the need for lightweight scaffolding before advice was given. For example, they suggested the use of short scales or check-in questions to assess severity and determine the appropriate depth or style of the response. Such scaffolding was not meant to copy formal clinical assessments, but to ensure interactions align with users’ current state, ease mental effort during moments of distress, and adjust the support they receive accordingly. This design opportunity presented by the participants is in line with previous work in HCI research that notes accessibility should be understood not as achieving perfect understanding, but as a process of ongoing attunement between systems and users ~\cite{Bennett2019}

\textbf{Integration Into Broader Care Systems} Participants did not view LLMs as stand-alone support, but as tools to be embedded within wider care ecosystems. They envisioned connections between everyday self-management and professional treatment, such as linking LLMs with data from wearables to provide more contextualized advice: ``so it gets that broader picture of you, your health in its broadest sense''~(P1) or enabling continuity with therapists by sharing conversation histories. In these accounts, the value of LLMs lies less in their isolated capabilities and more in how they can complement, extend, and bridge existing forms of care. Such integration highlights the opportunities to design LLMs not as replacements for therapists but as socio-technical systems. This vision provides empirical support for recent proposals to integrate LLMs into psychotherapy ~\citep{Stade2024LLMBehavioralHealthcare} and reinforces prior work arguing that trust and accountability require embedding AI into clinical workflows, not deploying it in isolation ~\cite{blackbox2020}.

\subsection{Limitations}

Although our sample was almost equally divided between male and female participants and included a range of mental health conditions, it does not fully represent the broader population of individuals with mental health issues. Additionally, all participants had previous therapy experience, which potentially affected how they set boundaries with the use of LLMs and compared AI support to human care. This study was based on self-reported mental health conditions. While we applied additional filters through Prolific and survey responses to improve data quality, the self-reports may not perfectly correspond with clinical assessments. This research is centered on residents of the UK, where engagement with LLMs is shaped by local healthcare practices and waiting periods. Future research should delve into cross-cultural comparisons to deepen our understanding of how LLMs are integrated into self-management strategies. While there are some limitations, our findings provide valuable insights into AI adoption in mental health care by highlighting how users safely incorporate LLMs into their mental health management and actively define the boundaries of LLM appropriateness.

\section{Conclusions}

This study examined how people with mental health challenges engage with LLMs for support. Through 20 semi-structured interviews, we highlighted lived experiences that illuminate both the motivations for use and the limitations of these systems. Participants valued LLMs not only for their immediacy and non-judgmental stance, but also for the ability to engage in self-paced disclosure, to reframe and structure thoughts, and to experience a sense of relational engagement. Simultaneously, they drew clear boundaries by emphasizing that LLMs were inadequate for crises, trauma, or socially and emotionally complex situations that require human empathy and judgment. Drawing from these findings, we propose design and governance strategies that position LLMs as supplementary tools within broader care systems rather than as substitutes for therapists. By highlighting how users define the role of LLMs, this study contributes to the development of responsible, user-focused AI applications in mental health care.


\bibliographystyle{ACM-Reference-Format}
\bibliography{aimentalhealth}

\end{document}